\documentclass[revtex,preprint,groupedaddress,preprintnumbers,showpacs,amsmath,amssymb]{revtex4-1}

\usepackage{graphicx}%
\usepackage{dcolumn}%
\usepackage{bm}%

\usepackage{color}
\begin{document}
\title{Ion acceleration from laser-driven electrostatic shocks}
\affiliation{GoLP - Instituto de Plasmas e Fus\~ao Nuclear - Laborat\'orio Associado, Instituto Superior T\'ecnico, 1049-001 Lisbon, Portugal}
\author{F. Fiuza \footnote[1]{current address: Lawrence Livermore National Laboratory, Livermore, CA 94551, USA}, A. Stockem, E. Boella \footnote[2]{also at Dipartimento Energia, Politecnico di Torino, 10129 Turin, Italy}, R. A. Fonseca  \footnote[3]{also at DCTI/ISCTE Instituto Universitario de Lisboa, 1649-026 Lisbon, Portugal}, and L. O. Silva}
\affiliation{GoLP - Instituto de Plasmas e Fus\~ao Nuclear - Laborat\'orio Associado, Instituto Superior T\'ecnico, 1049-001 Lisbon, Portugal}
\author{D. Haberberger, S. Tochitsky, W. B. Mori, and C. Joshi}
\affiliation{Department of Electrical Engineering, University of California, Los Angeles, CA 90095, USA}

\begin{abstract} 
Multi-dimensional particle-in-cell simulations are used to study the generation of electrostatic shocks in plasma and the reflection of background ions to produce high-quality and high-energy ion beams. Electrostatic shocks are driven by the interaction of two plasmas with different density and/or relative drift velocity. The energy and number of ions reflected by the shock increase with increasing density ratio and relative drift velocity between the two interacting plasmas. It is shown that the interaction of intense lasers with tailored near-critical density plasmas allows for the efficient heating of the plasma electrons and steepening of the plasma profile at the critical density interface, leading to the generation of high-velocity shock structures and high-energy ion beams. Our results indicate that high-quality 200 MeV shock-accelerated ion beams required for medical applications may be obtained with current laser systems.
\end{abstract} 
\maketitle

\section{Introduction}
\label{sec1}

Collisionless shocks are pervasive in space and astrophysical plasmas, from the Earth's bow shock to Gamma Ray Bursters, and are known to be efficient particle accelerators \cite{bib:blandford, bib:jones}, even though the details of the acceleration physics are not yet fully understood. The fast progress in laser technology is bringing the study of near-relativistic collisionless shocks into the realm of laboratory plasmas. Intense ($I > 10^{18} ~\mathrm{Wcm}^{-2}$) laser-plasma interactions allow for efficient heating and compression of matter \cite{bib:wilks} and for the generation of relativistic flows relevant to the study of astrophysical collisionless shocks \cite{bib:fiuza}.

Apart from the importance of a better understanding of the fundamental physics associated with the formation of collisionless shocks, there has been a growing interest in exploring laser-driven shocks as compact particle accelerators \cite{bib:denavit, bib:silva, bib:wei, bib:habara, bib:d'humieres}. Electrostatic shocks can act as a ``moving wall" as they propagate through the plasma, reflecting background ions to up to twice the shock velocity due to the strong electric field associated with the shock front. Previous numerical studies of laser-driven electrostatic shocks have shown that the interplay between shock acceleration and target normal sheath acceleration (TNSA \cite{bib:mora}), can lead to the generation of energetic ions with a broad spectrum \cite{bib:denavit, bib:silva, bib:d'humieres}.

Energetic ion beams from compact laser-produced plasmas have potential applications in many fields of science and medicine, such as radiotherapy \cite{bib:bulanov, bib:malka}, isotope generation for medical applications \cite{bib:spencer}, proton radiography \cite{bib:borghesi}, and fast ignition of fusion targets \cite{bib:roth}. However, producing focusable, narrow energy spread, energetic beams has proved to be challenging. In particular, radiotherapy requires energy spreads of 1-10\% FWHM and beam energies in the range of $100-300$ MeV/a.m.u. \cite{bib:linz}. 

Recent experimental \cite{bib:haberberger} and numerical \cite{bib:fiuza2} results have shown the possibility of using tailored near-critical density plasmas to control the sheath fields at the rear side of the plasma and generate shock-accelerated, high-quality ion beams. An exponentially decreasing plasma profile at the rear side of the target leads to a uniform and low-amplitude sheath electric field from the expansion of hot electrons into vacuum \cite{bib:grismayer}. The slowly expanding ions are then reflected by the high-velocity shock formed as a result of the laser-plasma interaction, leading to the formation of a energetic beam with narrow energy spread \cite{bib:fiuza2}.

In this paper, we expand these recent results \cite{bib:fiuza2} by analyzing in detail the different plasma conditions that lead to the formation of electrostatic shocks in plasma and their influence in the properties of the reflected ion beams. We consider both the case of idealized semi-infinite plasmas with arbitrary density, temperature, and velocity, and the case of laser-driven near-critical density laboratory plasmas. We show that electrostatic shocks can be formed in strongly heated plasmas by the interaction of two regions of different density and/or different velocity, and that ion reflection will occur either for large density ratios or for a limited range of relative drift velocities. We then focus on the possibility of driving electrostatic shocks in near-critical density plasmas. We show that there is an interplay between different physical mechanisms associated with the laser-plasma interaction at near-critical density, including laser filamentation, electron heating, and density steepening. The setting up of a fast return current in thin targets is critical to heating the entire plasma volume and density steepening plays an important role in launching a shock capable of reflecting the slowly expanding background ions. The importance of the plasma scale length at the rear side of the target in order to control the quality of the accelerated ion beam is also demonstrated. Under optimal conditions, it is shown that this scheme is scalable to the production of high-quality (energy spread of $\sim 10\%$ FWHM) 100-300 MeV ion beams for medical applications with currently available laser systems.

This paper is organized as follows. In Section \ref{sec2}, we analyze the formation of electrostatic shocks and the characteristics of the accelerated ions from the interaction of plasmas with different temperatures, densities, and/or flow velocities. We first review the theory of shock formation and ion acceleration and then use particle-in-cell (PIC) simulations to study the properties of the shock and reflected ions as a function of the initial conditions and we discuss the possibility of controlling the quality of the accelerated ion beam. In Section \ref{sec3}, we study the possibility of reaching the required conditions for shock formation and high-quality ion acceleration in the laboratory from the interaction of moderately intense lasers with tailored near-critical density plasmas. We identify the important mechanisms that lead to the formation of a strong shock capable of reflecting background ions and we derive the optimal conditions for the generation of high-quality ion beams in laboratory, which are validated by multi-dimensional PIC simulations. We explore the scaling of the ion energy with laser intensity showing the possibility of generating 200 MeV protons required for radiotherapy with current laser systems. Finally, in Section \ref{sec4}, we summarize our results.


\section{Electrostatic shocks in plasmas}
\label{sec2}

The interpenetration of collisionless plasmas of different density, temperature, or velocity, leads to a wide range of instabilities and to the formation of nonlinear structures capable of trapping and accelerating charged particles. Depending on the exact nature of the instabilities that mediate these nonlinear structures, different dissipation mechanisms can occur and lead to the formation of shockwaves. Electrostatic shocks are typically associated with the excitation of ion acoustic waves (IAW) in plasmas with cold ions and high electron temperatures. As these waves grow, they start trapping particles, reaching high field amplitudes and leading to the formation of a shockwave. If the electrostatic potential energy associated with the shock front is higher than the kinetic energy of the upstream ions, these shockwaves can reflect the upstream ions to twice the shock velocity acting as an efficient ion accelerator.

\subsection{Theory}

To study the formation of electrostatic shocks, we consider the interaction of two adjacent plasma slabs with an electron temperature ratio of $\Theta = T_{e\,1}/T_{e\,0}$ and a density ratio of $\Gamma = N_{e\,1}/N_{e\,0}$. Electrostatic shock structures can be generated as a result of the expansion of plasma 1 (downstream) into plasma 0 (upstream). Here, electrostatic instabilities at the edge of the plasmas can develop leading to the build up of the potential at the contact discontinuity. Electrostatic shocks can be formed \cite{bib:shamel,bib:sorasio} as dissipation is provided by the population of trapped particles behind the shock and, for strong shocks, by the ion reflection from the shock front \cite{bib:sagdeev}. Kinetic theory can be used to describe such a system, where both free and trapped electron populations are taken into account. The ions are treated as a fluid. The kinetic theory for the scenario, whereby an electrostatic shock is supported by regions/slabs of arbitrary temperature and density ratios has been outlined by Sorasio \emph{et. al.} \cite{bib:sorasio} to study the formation of high Mach number shocks.

\begin{figure}[!t]
\begin{center}
\includegraphics[width=1.\columnwidth]{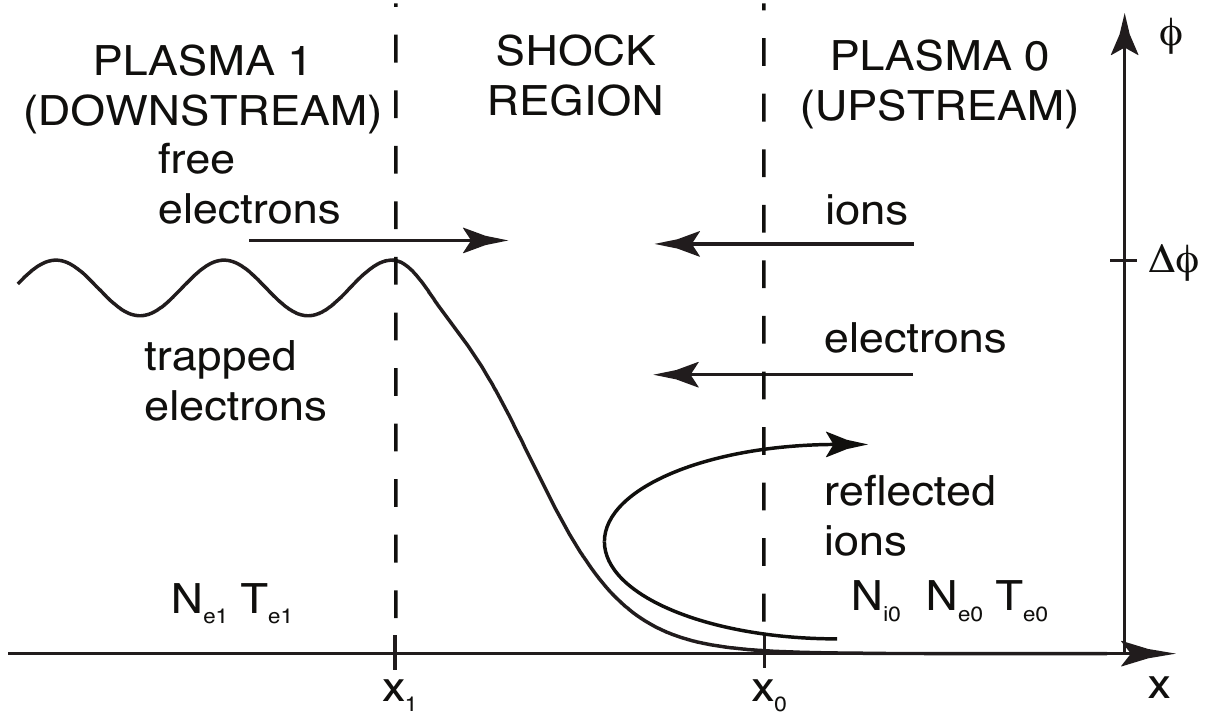}
\caption{\label{fig:theory_scheme} Steady state electrostatic shock structure as seen from the shock frame. Electrons from the upstream region move freely, while electrons from the downstream region can be either free or trapped. Ions, which flow from upstream to downstream, are slowed down by the electrostatic potential, and reflected back into the upstream for strong shocks.}
\end{center}
\end{figure}

The shock transition region is modeled in the reference frame of the shock; the electrostatic potential increases monotonically from $\phi = 0$ at $x = x_0$ to $\phi = \Delta \phi$ at $x = x_1$, as illustrated in Figure \ref{fig:theory_scheme}. The electron distribution $f_e (x, v_e)$ must be a solution of the stationary Vlasov equation and can be determined, at a given position $x$, as a function of the distribution of the plasma at the left ($x_1$) and right ($x_0$) boundaries. The free electron population propagating from the upstream to the downstream region is described by a drifting Maxwell-Boltzmann (MB) distribution function, with temperature $T_{e\,0}$ and fluid velocity $v_{sh}$ (in the laboratory frame, the upstream is assumed to be stationary), $f_{ef\,0} (v_0) = \frac{2 N_{e\,0}}{v_{th\,0}\sqrt{2 \pi}} e^{-\frac{(v_0 - v_{sh})^2}{2 v_{th\,0}^2}}$, where $N_{e\,0}$ is the density of electrons moving from the upstream to the downstream region and $v_{th\,0}$ is their thermal velocity, defined as $v_{th\,\alpha} = \sqrt{k_B T_{e\,\alpha}/m_e}$, with $k_B$ being the Boltzmann constant and $m_e$ the electron mass. We assume that the difference between the downstream velocity and the shock velocity is much smaller than the electron thermal velocity and, therefore, that the fluid velocity of the free electrons in the downstream region is approximately equal to zero in the shock frame. The free electrons in the downstream region have a MB distribution $f_{ef\,1} (v_1) = \frac{2 N_{e\,1}}{v_{th\,1}\sqrt{2 \pi}} e^{-\frac{v_1^2}{2 v_{th\,1}^2} + \frac{e \Delta \phi}{k_B T_{e\,1}}}$, where $N_{e\,1}$ is the density of electrons moving from the downstream to the upstream region and $v_{th\,1}$ is their thermal velocity. The trapped electron population is represented by a flat-top distribution function $f_{et\,1} = \frac{2 N_{e\,1}}{v_{th\,1}\sqrt{2 \pi}}$, following the maximum-density-trapping approximation \cite{bib:shamel}, which guarantees $f_{ef\,1} (v_1 = v_c) = f_{et\,1}$ at the critical velocity $v_c = \sqrt{\frac{2e \Delta \phi}{m_e}}$ that discriminates between free ($v_1 < - v_c$) and trapped electrons ($|v_1| < v_c$). The electron velocity at a given point follows from energy conservation $v_e = \sqrt{v_0^2 + \frac{2e\phi}{m_e}} = - \sqrt{v_1^2 + \frac{2 e (\phi-\Delta \phi)}{m_e}}$. The electron density along the shock transition is calculated by integrating the electron distribution function, yielding $n_0 (\varphi) = N_{e\,0} e^\varphi \mathrm{Erfc}[\sqrt{\varphi}]$ for electrons flowing from the upstream to the downstream and $n_1 (\varphi) = N_{e\,1} \Gamma e^{\varphi/\Theta} \mathrm{Erfc}[\sqrt{\varphi/\Theta}] + \frac{4}{\sqrt{\pi}} N_{e\,0} \Gamma \sqrt{\varphi/\Theta}$ for electrons flowing in the opposite direction, where $\varphi = \frac{e\phi}{k_B T_{e\,0}}$ and Erfc is the complementary error function. The ion density is determined from energy and mass conservation, yielding $n_i = N_{i\,0}/\sqrt{1-2\varphi/M^2}$, where $M = v_{sh}/c_{s\,0}$ is the shock Mach number, $c_{s\,0} = (k_B T_{e0}/m_i)^{1/2}$ is the upstream sound speed, and $m_i$ and $m_e$ are the ion and electron mass. Using charge neutrality at $x = x_0$ we obtain $N_{e\,0} = N_{i\,0} = N_0$.

The ion and electron densities can be combined with Poisson's equation to find the evolution of the electrostatic potential, which is given by $\frac{1}{2}\left(\frac{\partial \varphi}{\partial \chi}\right)^2 + \Psi(\varphi) = 0$, where $\chi = x/\lambda_D$, $\lambda_D = \sqrt{k_B T_{e\,0}/4 \pi e^2 N_0}$ is the Debye length, and the nonlinear Sagdeev potential \cite{bib:sagdeev} is given by
\begin{equation}
\Psi (\varphi) = P_i (\varphi, M) - P_{e\,1} (\varphi, \Theta, \Gamma) - P_{e\,0} (\varphi, \Gamma),
\end{equation}
where $P_{e\,1} (\varphi, \Theta, \Gamma) = P_{e\,f\,1} (\varphi, \Theta, \Gamma) + P_{e\,t\,1} (\varphi, \Theta, \Gamma) = \Theta \Gamma/(1+\Gamma)(e^{\varphi/\Theta} \mathrm{Erfc} \sqrt{\varphi/\Theta} + 2 \sqrt{\varphi/\pi \Theta} + (8/3) \varphi^{3/2}/\sqrt{\pi \Theta^3} -1)$ is the downstream electron pressure, $P_{e\,0} (\varphi, \Gamma) = 1/(1+\Gamma)(e^\varphi \mathrm{Erfc} \sqrt{\varphi} + 2 \sqrt{\varphi/\pi} -1)$ is the upstream electron pressure, and $P_i (\varphi, M) = M^2(1-\sqrt{1-2\varphi/M^2})$ is the ion pressure. The definition of $\varphi (x_0) = 0$ and the condition of charge neutrality at $x_0$ impose $\Psi (\varphi = 0) = 0$ and $\frac{\partial \Psi}{\partial \varphi}(\varphi = 0) = 0$, respectively. 

Shock solutions can be found for $\Psi (\varphi) < 0$, allowing for a complete description of the shock properties \cite{bib:tidman}. Ion reflection from the shock front will occur when the electrostatic potential across the shock exceeds the kinetic energy of the upstream ions, $e \phi > (1/2) m_i v_{sh}^2$, which corresponds to the critical value
\begin{equation}
\varphi_{cr} = \frac{M_{cr}^2}{2}.
\end{equation}
Although ion reflection is not included in this analysis, this critical condition can be used to infer the required shock properties, as a function of the plasma parameters, that lead to ion reflection from shocks. The critical Mach number, $M_{cr}$, for ion reflection can be found by solving numerically
\begin{equation}
M_{cr}^2 = \frac{\frac{\sqrt{2}M_{cr}}{\sqrt{\pi}} + e^{\frac{M_{cr}^2}{2}} \mathrm{Erfc}\left[\frac{M_{cr}}{\sqrt{2}}\right] -1 + \Gamma \Theta \left(\frac{\sqrt{2} M_{cr}}{\sqrt{\pi \Theta}} + e^{\frac{M_{cr}^2}{2 \Theta}} \mathrm{Erfc}\left[\frac{M_{cr}}{\sqrt{2\Theta}}\right] + \frac{4 M_{cr}^3}{3\sqrt{2 \pi \Theta^3}} -1\right)}{1+ \Gamma}.
\label{eq:Mcr}
\end{equation}

In order to study shock formation and ion acceleration in plasmas where the electron temperature is relativistic, we generalize this framework to relativistic temperatures \cite{bib:stockem}. Electrons are described by relativistic Juttner distributions 
\begin{equation} 
f_{ef\,0} (\gamma_0) = \frac{N_{e\,0}}{K_1[\mu_{e\,0}]}\frac{\gamma_0}{\sqrt{\gamma_0^2-1}} e^{-\mu_{e\,0} \gamma_0},
\label{eq:f0}
\end{equation}
\begin{equation} 
f_{ef\,1} (\gamma_1) = \frac{N_{e\,1}}{K_1[\frac{\mu_{e\,0}}{\Theta}]} \frac{\gamma_1}{\sqrt{\gamma_1^2-1}} e^{-\frac{\mu_{e\,0}}{\Theta} \gamma_1 + \frac{\varphi}{\Theta}},
\label{eq:f1}
\end{equation}
\begin{equation} 
f_{et\,1} =  \frac{N_{e\,1} e^{-\frac{\mu_{e\,0}}{\Theta}}}{K_1[\frac{\mu_{e\,0}}{\Theta}]} \frac{\gamma_1}{\sqrt{\gamma_1^2-1}},
\label{eq:t1}
\end{equation}
where $\gamma_{0,1}$ are the relativistic Lorentz factors of upstream and downstream electrons, respectively, $\mu_{e\, 0} = m_e c^2/k_B T_{e\, 0}$, and $K_1$ is the modified Bessel function of the second kind.

The generalized electron pressures are found by following the same procedure as for the non relativistic case and are given by 
\begin{equation} 
P_{e\,0} (\varphi, \Gamma, \mu_{e\,0}) = \frac{1}{1+\Gamma} \left[ \left (\frac{\mu_{e\,0}}{K_1 [\mu_{e\,0}]} \int_1^\infty d\gamma e^{-\mu_{e\,0} \gamma} \sqrt{\left(\gamma+\frac{\varphi}{\mu_{e\,0}}\right)^2-1}\right) -1\right],
\label{eq:p0}
\end{equation}
\begin{equation} 
P_{e\,f\,1} (\varphi, \Theta, \Gamma, \mu_{e\,0}) = \frac{\Gamma \Theta}{1+\Gamma} \left[\left(\frac{\mu_{e\,0}}{\Theta K_1 [\mu_{e\,0}/\Theta]} \int_1^\infty d\gamma e^{- \frac{\mu_{e\,0} \gamma}{\Theta}} \sqrt{\left(\gamma+\frac{\varphi}{\mu_{e\,0}}\right)^2-1} \right)-1\right],
\label{eq:pf1}
\end{equation}
\begin{equation} 
P_{e\,t\,1} (\varphi, \Theta, \Gamma, \mu_{e\,0}) = \frac{\Gamma}{1+\Gamma}  \frac{\mu_{e\,0} e^{-\frac{\mu_{e\,0}}{\Theta}}}{ K_1 [\mu_{e\,0}/\Theta]} \left(\sigma \sqrt{\sigma^2 -1} - \mathrm{Log} [\sigma + \sqrt{\sigma^2-1}] \right),
\label{eq:pt1}
\end{equation}
where $\sigma = 1+\varphi/\mu_{e\,0}$. In the relativistic limit, $\mu_{e\, 0} \ll 1$, and we get $P_{e\,1} (\varphi, \Theta, \Gamma, \mu_{e\,0}) = \varphi \Gamma[\mu_{e\,0}(1-\varphi/\Theta) + \varphi + \Theta]/[(1+\Gamma)\Theta]$ and $P_{e\,0} (\varphi, \Theta, \Gamma, \mu_{e\,0}) = \varphi (1-\mu_{e\,0})/(1+\Gamma)$. The critical Mach number is given by
\begin{equation} 
M_{cr} = \sqrt{2 \Theta \left( \frac{1+ \mu_{e\,0}}{\Gamma (1- \mu_{e\,0}/\Theta )} +1\right)}.
\label{eq:mcrit}
\end{equation}

In the limit of large density ratios ($\Gamma \gg 1$), $M_{cr} = \sqrt{2 \Theta}$ and in the limit of low density ratios ($\Gamma \ll 1$), $M_{cr} \propto \sqrt{\Theta/\Gamma}$. Ion reflection can therefore occur for low/moderate Mach number shocks provided that $\Gamma \gg 1$ and $\Theta \sim 1$.

\subsection{Shock formation}

In order to validate the theoretical predictions for the electrostatic shock structure and the conditions for ion reflection we have performed 2D OSIRIS \cite{bib:fonseca} simulations of the interaction of two plasmas with different densities, temperatures, and relative velocity. Full-PIC simulations allow us to understand in a detailed and fully self-consistent way the formation of the shock structure and the properties of the reflected ions, as they capture the different kinetic processes involved.

We model the interaction of two semi-infinite plasmas and we vary their initial relative temperature, density, and/or drift velocity. We consider plasmas with non-relativistic (1 keV) and relativistic (1.5 MeV) electron temperatures. We use a simulation box with $4098 \times 128 (c/\omega_{p1})^2$, where $\omega_{p1} = \sqrt{4 \pi n_1 e^2/m_e}$ is the electron plasma frequency of the denser plasma (slab 1), which is located on the left-hand side of the simulation box, between $x_1 = 0$ and $x_1 = 2048 c/\omega_{p1}$. The plasma slab 0 is located between $x_1 = 2048 c/\omega_{p1}$ and $x_1 = 4096 c/\omega_{p1}$, and, therefore, the contact point of the two slabs is at $x_1 = 2048 c/\omega_{p1}$. In simulations with different density, temperature, and/or drift velocity between the two slabs, slab 1 is always the slab with higher density, temperature, and/or drift velocity, and will correspond to the downstream plasma once a shock is formed. Slab 0 thus corresponds to the upstream plasma region. The size of the numerical grid is chosen in order to resolve the smallest of the relevant plasma scales (either the Debye length or the electron skin depth) with at least 2 points in each direction. For instance, for $T_e = 1.5$ MeV, $\Delta x_1 = \Delta x_2 = 0.5 c/\omega_{p1} = 0.3 \lambda_D$ and $\Delta t = 0.3 \omega_{p1}^{-1}$. We use $9 - 36$ particles per cell per species together with cubic particle shapes and current smoothing for good accuracy.

\begin{figure}[!t]
\begin{center}
\includegraphics[width=1.\columnwidth]{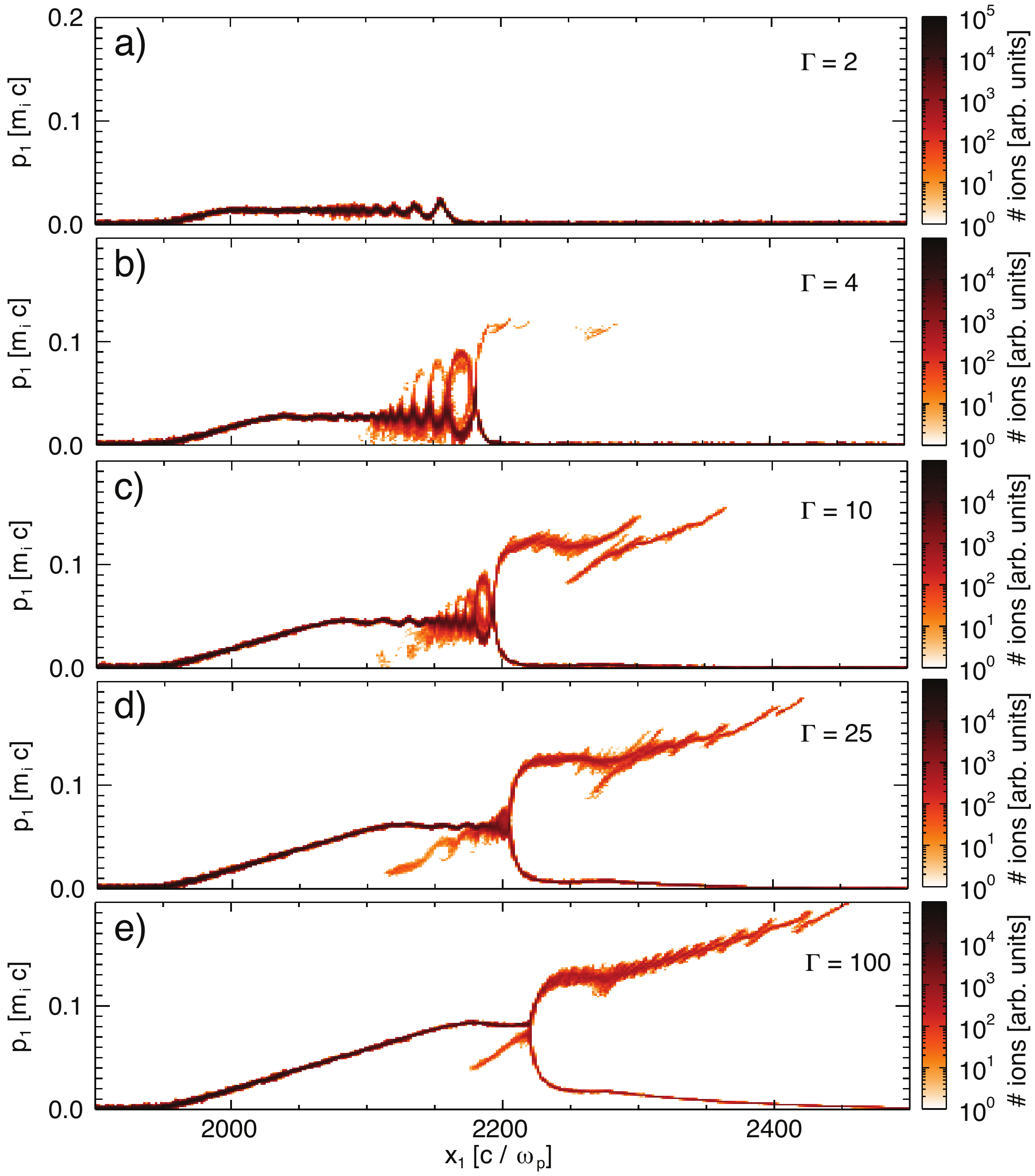}
\caption{\label{fig:p1x1_n} Ion phase space structure as a function of the initial density ratio $\Gamma$ between two plasma slabs/regions for $\Theta = 1$ and $T_e = 1.5$ MeV. Snapshots are taken at $t = 2450 ~\omega_{p1}^{-1}$. At $t = 0$ there is no relative drift between the two slabs.}
\end{center}
\end{figure}

\begin{figure}[!t]
\begin{center}
\includegraphics[width=1.\columnwidth]{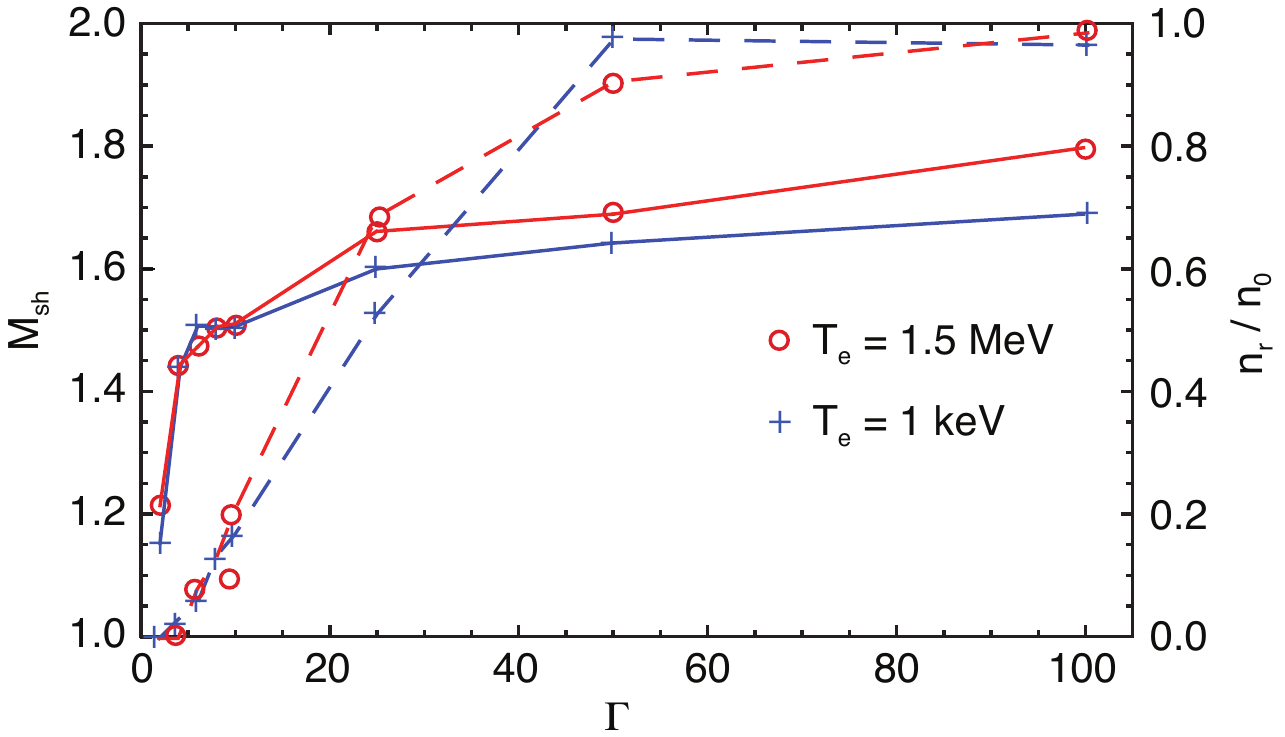}
\caption{\label{fig:M_n_r} Shock Mach number (solid lines) and fraction of ions reflected from the upstream (dashed lines) as a function of the initial density ratio $\Gamma$ between two plasma slabs/regions for $\Theta = 1$ and  $v_{1,0} = 0$ .}
\end{center}
\end{figure}

Figure \ref{fig:p1x1_n} illustrates the ion phase space for different initial density ratios $\Gamma = 2 - 100$ between the two plasma slabs. For very small density ratios ($\Gamma \simeq 2$) the expansion of the denser plasma into the more rarefied one drives a nonlinear IAW but no ion reflection is present (Fig. \ref{fig:p1x1_n} a). As $\Gamma$ increases, the amplitude of the IAW increases and ion trapping becomes evident. Around $\Gamma = 4$, the electrostatic field associated with the leading edge of the IAW gets high enough to start reflecting ions from the background plasma (Fig. \ref{fig:p1x1_n} b). For very high $\Gamma$, ion reflection becomes dominant, with the majority of the upstream ions being reflected by the shock structure and the trapped component becomes less noticeable (Fig. \ref{fig:p1x1_n} c-e). Both the shock Mach number $M_{sh}$ and the fraction of upstream ions reflected by the shock $n_r/n_0$ increase with the density ratio $\Gamma$ as shown in Figure \ref{fig:M_n_r}. For a plasma with initial constant electron temperature ($\Theta = 1$) and no drift velocity, the maximum Mach number is observed to be between 1.6 and 1.8.

\begin{figure}[!t]
\begin{center}
\includegraphics[width=1.\columnwidth]{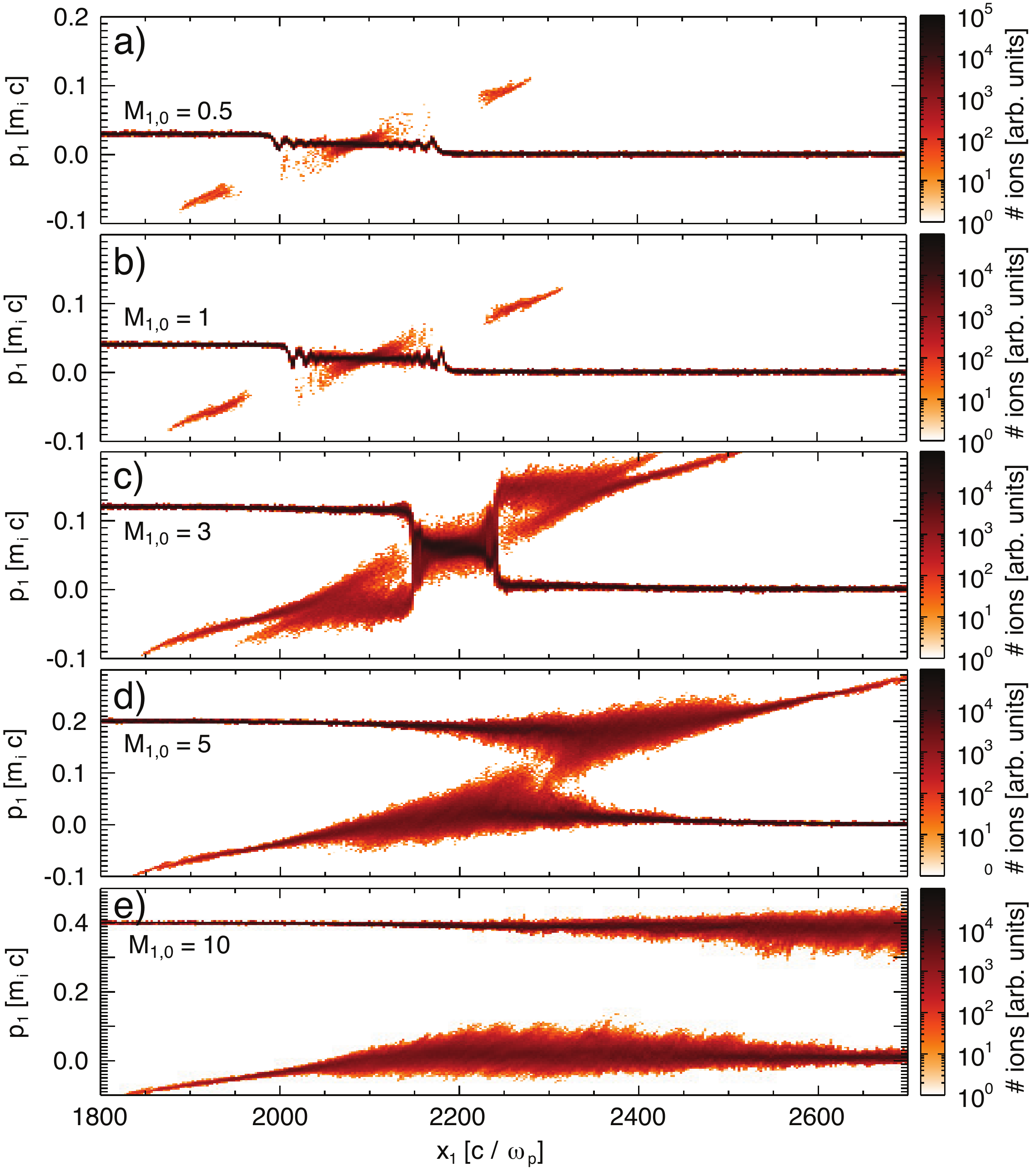}
\caption{\label{fig:p1x1_M} Ion phase space structure as a function of the initial relative drift between two plasma slabs/regions for $\Gamma = \Theta = 1$. Snapshots are taken at $t = 2450 ~\omega_p^{-1}$.}
\end{center}
\end{figure}

\begin{figure}[!t]
\begin{center}
\includegraphics[width=1.\columnwidth]{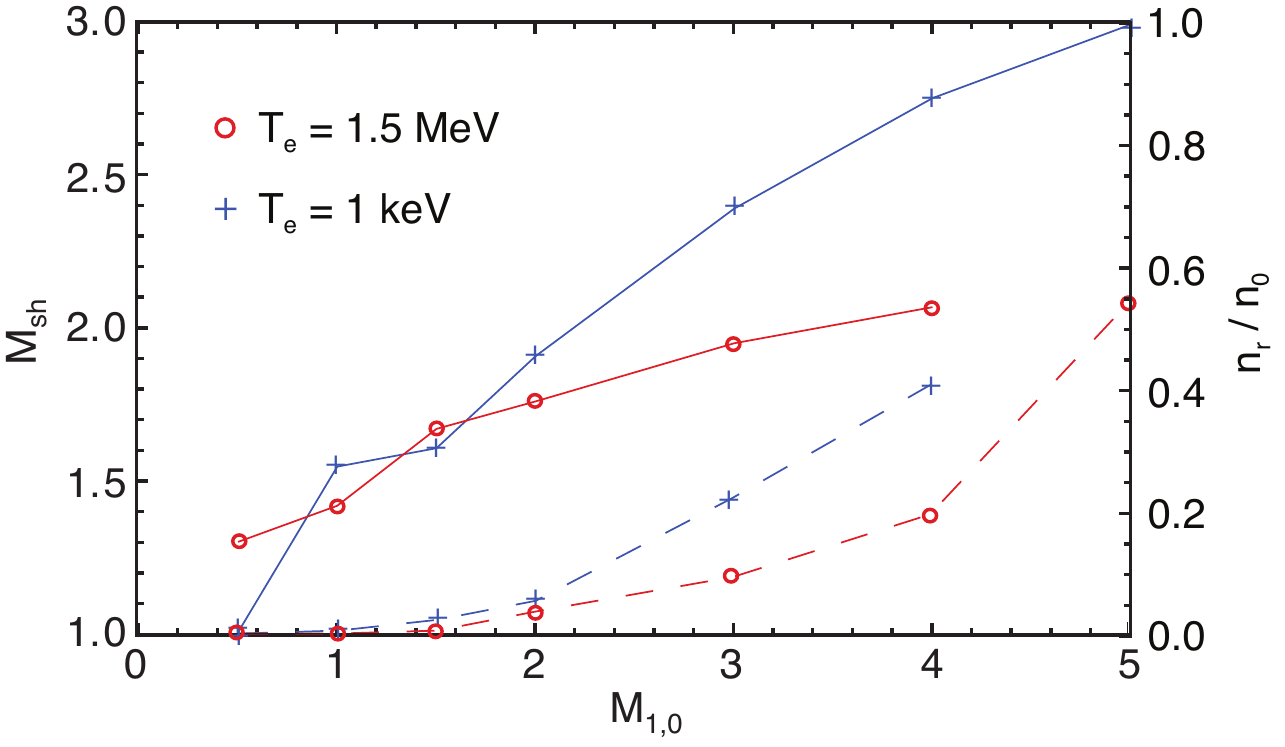}
\caption{\label{fig:M_v_r} Shock Mach number (solid lines) and fraction of ions reflected from the upstream (dashed lines) as a function of the initial Mach number of the relative drift between two plasma slabs/regions for $\Gamma = \Theta = 1$.}
\end{center}
\end{figure}

We have also studied the influence of an initial relative drift between the two plasma slabs for $\Gamma = \Theta = 1$ (see Fig. \ref{fig:p1x1_M}). For low relative drift velocities a nonlinear IAW is again formed but does not allow for significant particle trapping and no ion reflection is observed (Fig. \ref{fig:p1x1_M} a and b), as in the case of low $\Gamma$. As the relative drift velocity is increased, the amplitude of the IAW becomes larger and particles are trapped and reflected by the shock. The shock is observed to start reflecting ions for a relative Mach number between the two slabs $M_{1,0} = v_{1,0}/c_{s\,0} \sim 3$ (Fig. \ref{fig:p1x1_M} c). Again, both the Mach number of the generated shock and the fraction of reflected ions increase with the relative drift velocity between the two plasma slabs. For $\Gamma = 1$ and $\Theta = 1$, the maximum $M_{sh}$ reached is between $2-3$ for $M_{1,0} \simeq 4 - 5$ (Fig. \ref{fig:M_v_r}). For very large relative flows ($M_{1,0} > 10$), as the relative drift velocity starts approaching the electron thermal velocity, $v_{1,0} \approx v_{th}$, the kinetic energy of the flow is much larger than the electrostatic energy at the contact discontinuity and the flows are only weakly perturbed. For the simulated times ($t \leq 10^4 \omega_p^{-1}$) no shock is formed (Fig. \ref{fig:p1x1_M} d and e). In the opposite limit, when $v_{1,0} \gg v_{th}$, two-stream and Weibel-type instabilities \cite{bib:fiuza} are expected to dominate the shock formation process.

As the temperature ratio between the two slabs is increased, larger shock Mach numbers can be reached and a wider range of relative drift velocities can lead to the formation of electrostatic shocks. For instance, for $\Theta = 10$ and $M_{1,0} = 10$ a shock is formed with $M_{sh} = 7$ and for $\Theta = 100$ and $M_{1,0} \sim 20$ shock Mach numbers as high as 20 can be reached. The laboratory study of such high Mach numbers \cite{bib:kuramitsu} would provide important insight on the formation of electrostatic shocks in space with $M_{sh} = 20 - 100$. In simulations where the two plasma slabs have different temperatures but the same density and no initial relative velocity, no shock is expected and none has been observed.

\subsection{Ion acceleration}

\begin{figure}[!t]
\begin{center}
\includegraphics[width=1.\columnwidth]{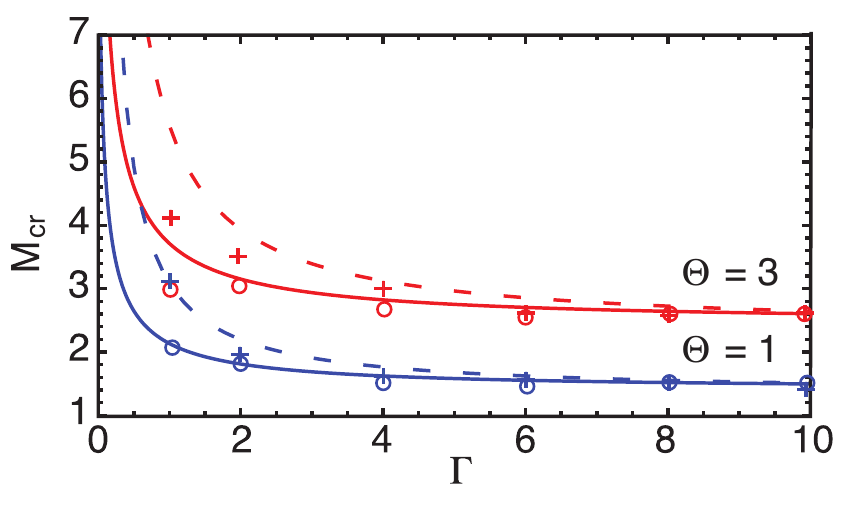}
\caption{\label{fig:theory} Critical Mach number for ion reflection in electrostatic shocks as a function of the density ratio $\Gamma$ and temperature ratio $\Theta$ between the two plasma slabs/regions, for $T_{e\,0} = 1$ keV (dashed line \cite{bib:sorasio}) and $T_{e\,0} = 1.5$ MeV (solid line Eq. \eqref{eq:mcrit}). The symbols indicate the simulation values for the non-relativistic (+) and relativistic (o) electron temperatures, which were obtained by measuring the speed of the shock structure (density jump or electrostatic field) when ion reflection is observed.}
\end{center}
\end{figure}

From the study of the formation of electrostatic shocks for different relative densities, temperatures, and drift velocities it is possible to infer the critical Mach number for ion reflection, $M_{cr}$. For a given combination of initial density ratio $\Gamma$ and temperature ratio $\Theta$, we have varied the initial drift velocity between the two plasma slabs in order to determine the lowest Mach number for which ion reflection is observed, which corresponds to $M_{cr}$. Figure \ref{fig:theory} illustrates $M_{cr}$ as a function of $\Gamma$ and $\Theta$. We observe that the critical Mach number for ion reflection is in good agreement with theory (Eqs. \eqref{eq:Mcr} and \eqref{eq:mcrit}), as indicated by the red and blue circles and crosses in Figure \ref{fig:theory}. At high density ratios $\Gamma \ge 4$, the expansion of the two slabs (initially at rest) is sufficient to form the shock and reflect the ions. At lower density ratios, the plasma slabs need to have an initial relative drift in order to reach $M_{cr}$ for ion reflection. The Mach numbers observed in PIC simulations when ion reflection occurs lie very near to the theoretical curve for $M_{cr}(\Gamma,\Theta)$ and therefore we can consider that the ion velocity will be given by $v_{ions} \propto 2 M_{cr} c_{s\,0}$. The acceleration of ions to high energies in the shock requires strong electron heating in order to increase the sound speed.

\begin{figure}[!t]
\begin{center}
\includegraphics[width=1.\columnwidth]{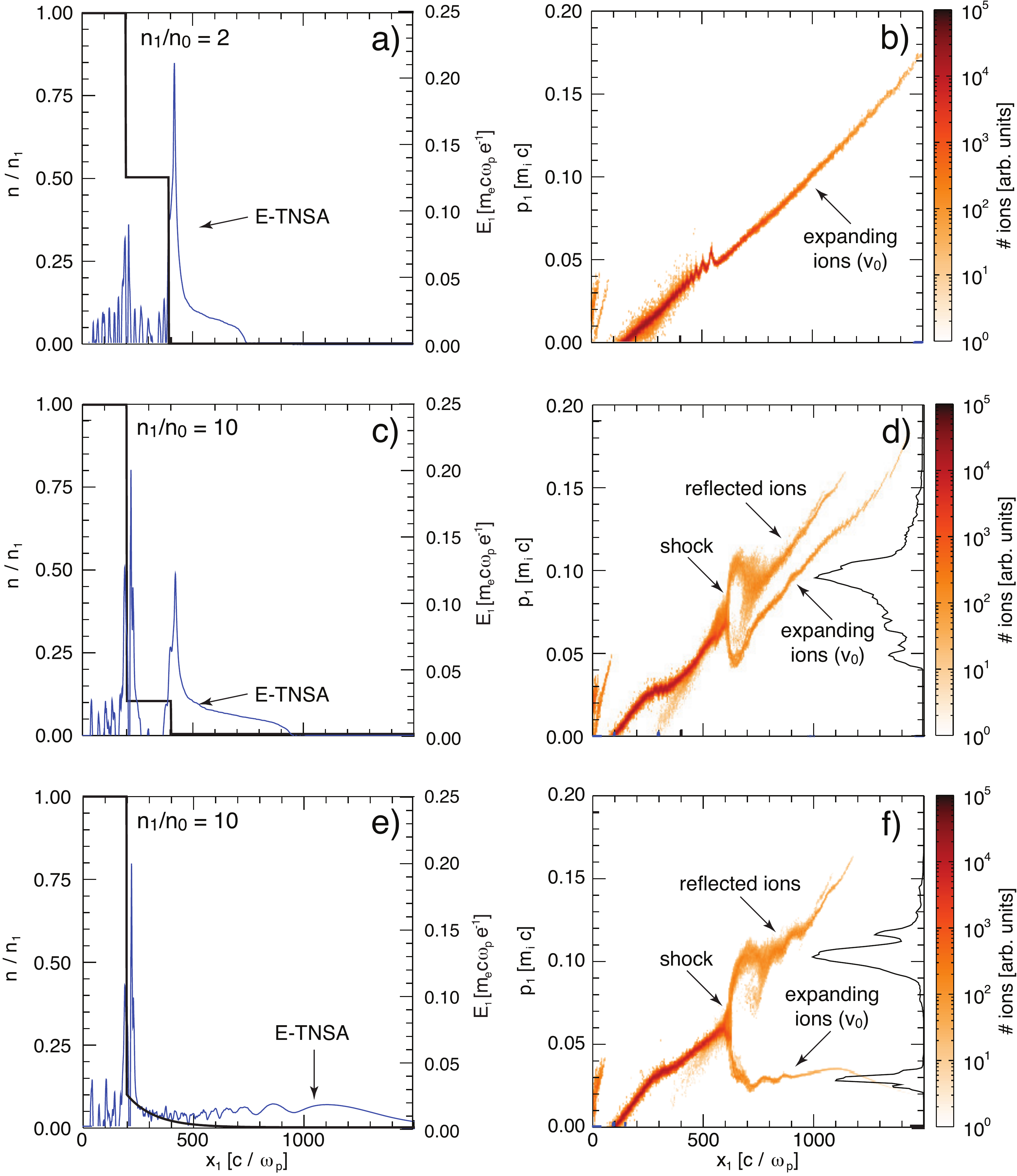}
\caption{\label{fig:slabs} Electric field structure and accelerated ion spectrum from the interaction of two finite plasma slabs with a density ratio (a,b) $\Gamma = 2$, (c,d) $\Gamma = 10$, and (e,f) $\Gamma = 10$ followed by an exponentially decreasing profile. Initially, $\Theta = 1$ ( $T_e = 1.5 MeV$) and $v_{1,0} = 0$. Left panels show the initial density profile (black) and early time longitudinal electric field (blue), whereas the right panels show the ion phase space (orange) and the spectrum of ions ahead of the shock (black line) at t =  $7700 ~\omega_{p1}^{-1}$.}
\end{center}
\end{figure}

In more realistic plasma configurations, where finite plasma slabs are considered, it is important to address the expansion of hot electrons into vacuum and the role of the resulting space-charge field on the quality of the shock accelerated ion beam. This TNSA field will accelerate the upstream ions to a given velocity $v_0$. The shock will then reflect the upstream ions to a velocity $v_{ions} \simeq 2 M_{cr} c_{s\,0} + v_0$. To investigate the role of competing fields in finite size plasmas we have preformed 2D simulations where each plasma slab has a thickness of $200 c/\omega_{p1}$ and are followed by a vacuum region. In the first case, we use a density ratio $\Gamma = 2$ (Figs. \ref{fig:slabs} a and b) and in the second case $\Gamma = 10$ (Figs. \ref{fig:slabs} c and d). In both cases $\Theta=1$ ($T_e = 1.5$ MeV). For the abrupt plasma-vacuum transition, the electrostatic field in the sheath at the rear side of the upstream plasma is nonuniform and introduces a chirp in $v_0$, broadening the ion energy spectrum as typical of TNSA \cite{bib:mora} (Figs. \ref{fig:slabs} b and d). This sheath field can be controlled by using an expanded plasma profile in the upstream slab. For an exponential plasma profile with scale length $L_g$, the sheath electric field is constant at early times ($t \ll 4 L_g/c_{s\,0}$) \cite{bib:grismayer} and its amplitude is given by
\begin{equation} 
E_{TNSA} = \frac{k_b T_{e\,0}}{e L_g}.
\label{eq:efield}
\end{equation}
A uniform sheath field will preserve the monoenergetic ion distribution as particles are reflected by the shock. This is illustrated in Figs. \ref{fig:slabs} e and f, where we replace the low density slab of Figure \ref{fig:slabs} a with an exponentially decreasing profile starting from the same peak density. The TNSA field is now approximately uniform (Fig. \ref{fig:slabs} e) leading to a slow expansion at uniform velocity of the upstream ions (Fig. \ref{fig:slabs} f). These expanding ions are then reflected by the electrostatic shock and cross the sheath region while preserving their narrow energy spread (Fig. \ref{fig:slabs} f), thus indicating a configuration suitable for the generation of monoenergetic ion beams.

These results indicate that high energy and high quality ion beams can be produced from shockwave acceleration in heated plasmas with an exponentially decreasing density profile. In order to achieve good quality in the accelerated ion beam it is necessary to guarantee that the velocity of the expanding upstream ions, $v_0$, is significantly smaller than the shock velocity by the time the shock is formed and starts reflecting the upstream ions, $\tau_r$, \emph{i.e.} $v_{sh} \gg c_{s\,0}^2 \tau_r / L_g$.  For strong shocks, where ion reflection is the dominant dissipation mechanism, the ion reflection time, $\tau_r$, is similar to the shock formation time and corresponds to the time an upstream ion takes to accelerate to $v_{sh}$ in the presence of the shock electrostatic field. Viewed another way, in the shock frame, where the upstream ions are moving towards the shock at $- v_{sh}$, reflection occurs when the electric field associated with the shock is able to stop the incoming ions. For the sake of simplicity, we assume the upstream ions initially at rest ($v_0 = 0$) and a uniform electric field, $E_{sh} = -\phi/L_{sh}$, within the shock transition region, $L_{sh}$, which for electrostatic shocks is of the order of the Debye length, $\lambda_D$. Let us use $L_{sh} = \delta \lambda_D$, with $\delta = O(1)$. The reflection time can then be estimated as $\tau_r = \delta m_i v_{sh} \lambda_D/(e \phi)$. As we have seen, for shocks driven by the interaction of two plasma regions with different densities and low or null initial relative drift velocity, the shock Mach number lies near $M_{cr}$ and therefore we can use $e \phi = (1/2)m_i v_{sh}^2$, yielding
\begin{equation} 
\tau_r = \frac{2 \delta M_{sh}}{\omega_{pi}}.
\label{eq:time}
\end{equation}
We note that the obtained expression for the ion reflection/shock formation time is consistent with the numerical results obtained by Forslund and Shonk \cite{bib:forslund}, where the shock formation time increases approximately linearly with the $M_{sh}$ before reaching the critical Mach number, a for $M_{sh} \sim 1.5$ the shock formation time is $4 \pi/\omega_{pi}$. The necessary condition for the generation of monoenergetic ion beams can then be written as $L_g \gg 2 L_{sh}$.

\section{Laser-driven electrostatic shocks}
\label{sec3}

The conditions required to drive strong electrostatic shocks and generate monoenergetic ion beams can be obtained in practice from the interaction of an intense laser pulse with plasma. The rear side exponential profile, similar to that shown in Figure \ref{fig:slabs} e, can be naturally formed by ionization/pre-heating of the target and consequent expansion, for instance due to the laser pre-pulse or an earlier laser pulse of lower intensity. Previous work on electrostatic shock formation from laser-plasma interactions focused mainly on laser-solid interactions \cite{bib:denavit,bib:silva}, where electron heating occurs at the vacuum-plasma surface and then rely on collisionless plasma processes to heat up the dense background plasma. In this case, very high laser intensities are required in order to heat the high density electrons to MeV temperatures. Here, we focus on the use of near-critical density plasmas, for which the laser can interact with a significant volume of the target and efficiently heat the electrons. 

\subsection{Laser-plasma interaction at near-critical density}

As an intense laser propagates in a plasma with density varying from undercritical to critical, $n_{cr}$, it will be partially absorbed by heating up the plasma electrons. Depending on the laser intensity, polarization, and target density, different absorption and particle acceleration mechanisms can occur. For instance, in the underdense region of the target the laser can undergo filamentation \cite{bib:max}, self-focusing \cite{bib:max,bib:mori}, and stimulated Raman scattering \cite{bib:forslund2}. As it reaches near-critical densities it will steepen the plasma profile locally \cite{bib:estabrook} and heat electrons due to a $\bf{J} \times \bf{B}$ mechanism \cite{bib:kruer,bib:may}.

Assuming that the laser interacts with the majority of the plasma electrons, the electron temperature, $\alpha k_B T_e = \epsilon_e$, can be roughly estimated by equating the plasma electron energy density to the absorbed laser energy density, $\alpha a_0 n_c L_{target} k_B T_e = \eta I \tau_{laser}$, where $\alpha$ is $3/2$ for non-relativistic plasmas and $3$ in the relativistic case, $\eta$ is the absorption efficiency, and the relativistically corrected critical density $a_0 n_c$ has been used, yielding
\begin{equation} 
T_e [\mathrm{MeV}] \simeq 0.078 \frac{\eta}{\alpha} a_0 \frac{\tau_{laser} [\mathrm{ps}]}{L_{target} [\mathrm{mm}]}.
\label{eq:temperature}
\end{equation}
For relativistic laser intensities, $a_0 > 1$, and steep density profiles at the laser-plasma interaction region, the temperature of accelerated electrons is expected to be close to ponderomotive \cite{bib:may,bib:wilks}, which leads to a similar dependence $T_e \propto a_0$. Under these conditions and for a typical target size $L_{target} < 1$ mm, laser pulses with picosecond duration can heat the plasma electrons to MeV temperatures, leading to high shock velocities and high reflected ion energies.

In order to drive an electrostatic shock, apart from providing the electron heating, it is necessary to create a sharp density variation and/or a relative drift velocity between different regions of the plasma as seen in Section \ref{sec2}. The plasma push and density steepening due to the radiation pressure can provide the required conditions. As the laser is stopped around the critical density and steepens the plasma profile, the heated electrons propagate through the back side of the target, where they find unperturbed plasma at a similar density, driving a return current that pulls the background electrons to the laser region where they are accelerated. Therefore, thin targets with peak density around the critical density allow for an efficient heating of the entire plasma. 

The initial build up of the return current together with the quick recirculation of the heated electrons due to the space-charge fields at the front and at the back of the target, will lead to a uniform temperature profile \cite{bib:silva, bib:mackinnon}, which is crucial in order to have a uniform shock velocity and a uniform ion reflection. Therefore, the target thickness, $L_{target}$, should be limited in order to guarantee that electrons can recirculate in the target before ion reflection occurs. For a ion reflection time $\tau_r = 4 \pi/\omega_{pi}$ (low Mach number shocks \cite{bib:forslund}), the limit on the maximum target thickness is given by $L_{\mathrm{target}} < 2 \pi c/\omega_{pi}$, or equivalently for critical density plasmas $L_{\mathrm{target}} < \lambda_0 (m_i/m_e)^{1/2}$. 

As noted in the Section \ref{sec2}, in order to control the strong space-charge fields and maintain a narrow energy spread, it is important to have a large scale length at the rear side of the target. Therefore, the optimal target thickness should be close to the maximum thickness for uniform heating. For a symmetric target expansion ($L_{target} \leq 2 L_g$), the optimal target scale length for uniform electron heating and ion reflection is then \cite{bib:fiuza2}
\begin{equation} 
L_{g\,0} \approx \frac{\lambda_0}{2} \left(\frac{m_i}{m_e}\right)^{1/2}.
\label{eq:optimal}
\end{equation}

\subsection{Shock formation and ion acceleration}

In order to explore the physics of laser-plasma interaction at near-critical density and to validate the proposed scheme for the generation high-velocity electrostatic shocks and high-quality ion beams, we have performed 2D OSIRIS simulations. In this case we use a larger simulation box in order to accommodate a vacuum region on the left hand side of the target, where the laser interacts with the plasma, and an extended vacuum region on the right hand side, where the plasma will expand and ions will be accelerated. The simulation box size is $3840 \times 240 ~(c/\omega_0)^2$ and is resolved with $12288 \times 768$ cells. 

We model the interaction of a Gaussian laser pulse with a duration of $1885 \omega_0^{-1}$ (FWHM), infinite spot size, and a normalized vector potential $a_0 = 2.5$ with a plasma with peak density of $2.5 n_c$. The pre-formed electron-proton plasma profile has a linear rise over $10 \lambda_0$ and falls exponentially with scale length $L_g = 20 \lambda_0$ (chosen according to Eq. \eqref{eq:optimal}). 

\begin{figure*}[t!]
\begin{center}
\includegraphics[width=1.\textwidth]{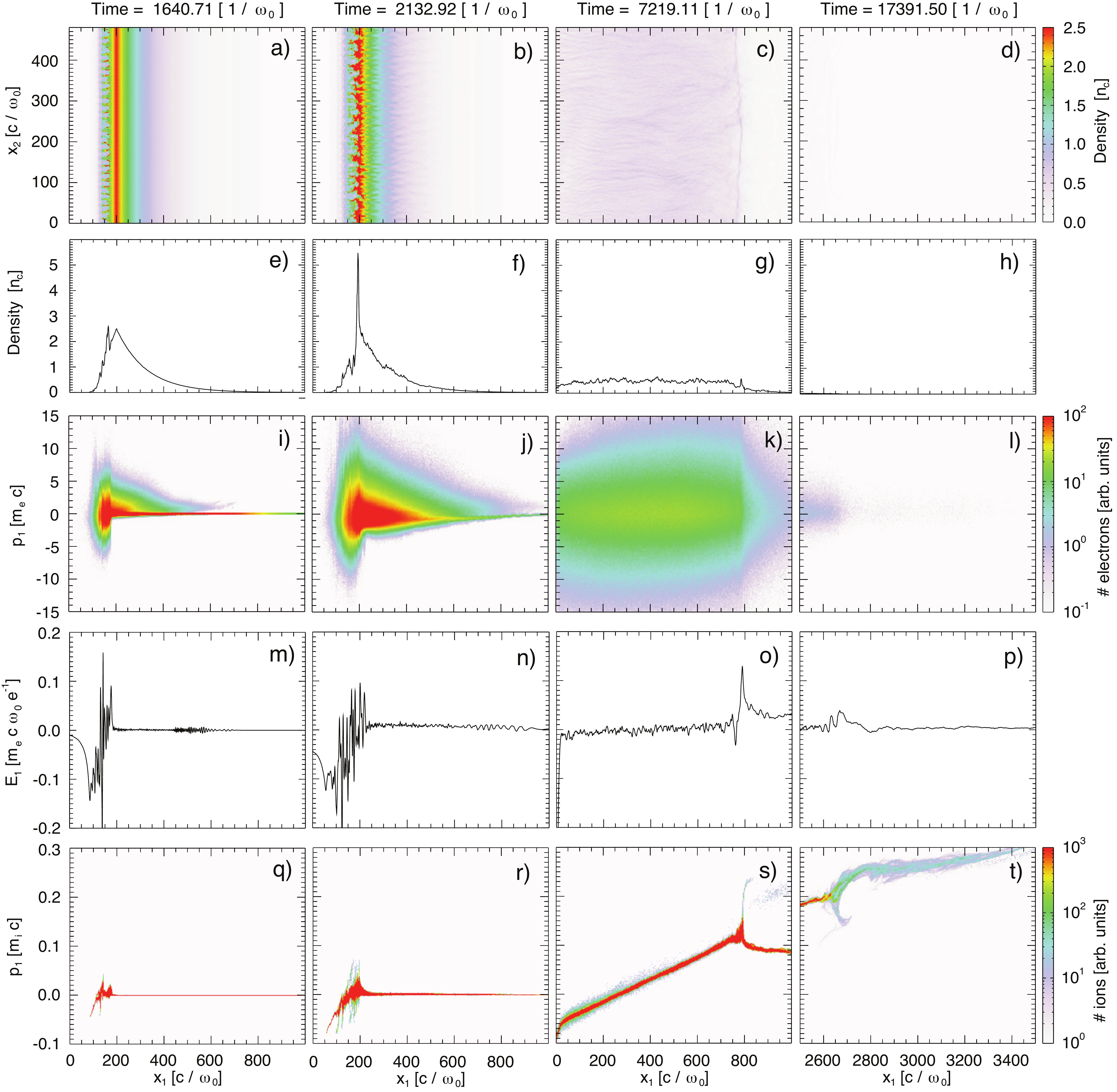}
\caption{\label{fig:shock} Temporal evolution of the laser-plasma interaction at near critical densities, from electron heating to shock formation, and ion acceleration. Row 1 shows the evolution of the ion density profile and row 2 shows a central lineout of the density along the laser propagation axis. Row 3 illustrates the evolution of the electron phase-space, row 4 the longitudinal electric field, and row 5 the ion phase-space.}
\end{center}
\end{figure*}

Figure \ref{fig:shock} illustrates the temporal evolution of the interaction. At early times, it is possible to observe the filamentation of the laser in the underdense plasma and strong electron heating (Fig. \ref{fig:shock} a and i). As the peak laser intensity reaches the critical density region, there is a clear steepening of the local density inside the filaments where the field is amplified. At this point, the peak density is increased by a factor of $2-4$ and followed by the exponential profile, similar to the case of Figure \ref{fig:slabs} e, which is critical for the shock to be driven. We note that the ions also gain a drift velocity at this critical density region due to the space-charge field caused by the electron acceleration. This drift velocity is measured to be $\sim 0.02 c$ (Fig. \ref{fig:shock} r), which is slightly smaller than the hole-boring velocity \cite{bib:wilks} $v_{hb} = a_0\sqrt{(n_{cr}/2n_p)(m_e/m_i)} = 0.026$ and corresponds to a Mach number of $\sim 0.4$ for the measured electron temperature, which is 2.2 MeV. Both the density jump and the drift velocity will contribute to the shock formation and the interplay between these two effects can be controlled by tuning the laser and plasma parameters. For the profile used, and taking into account the results obtained in Section \ref{sec2}, we expect the density jump to be the dominant effect in our case, and we observe an electrostatic shock being formed as the result of the expansion of the heated and tailored plasma profile (Fig. \ref{fig:shock} o). 

\begin{figure}[!t]
\begin{center}
\includegraphics[width=1.\columnwidth]{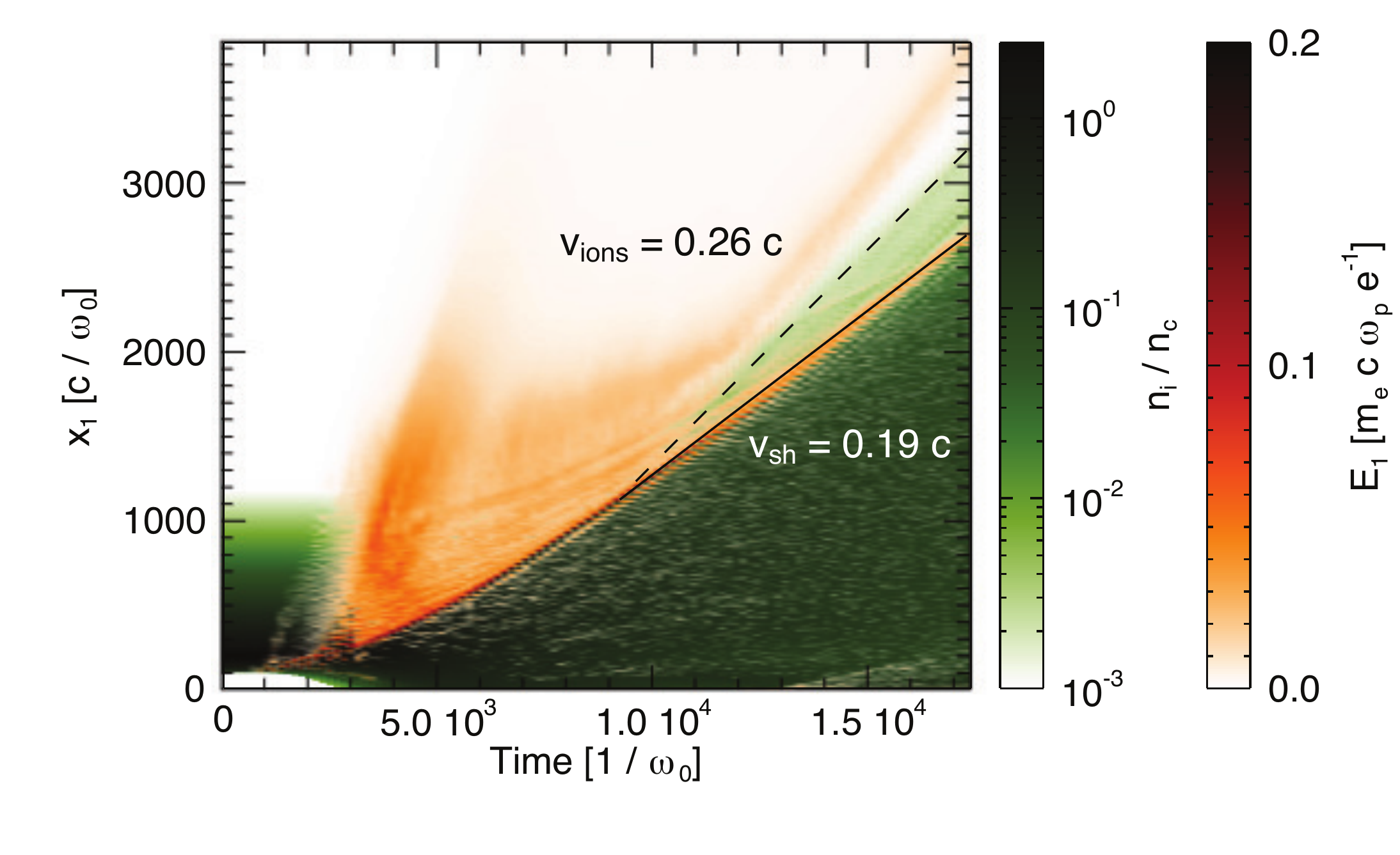}
\caption{\label{fig:x_t} Time evolution of the ion density (green) and longitudinal electric field (orange). The strong feature between $3\times10^3 \omega_0^{-1}$  and $4\times10^3 \omega_0^{-1}$ is associated with the laser plasma interaction and the fields driven by the fast electrons. The solid line follows the shock and the dotted line follows the reflected ions.}
\end{center}
\end{figure}

Although the majority of the laser light is stopped and cannot interact with the electrons at the rear side of the target, a return current is set up due to the current imbalance produced by the fast electrons in the unperturbed plasma. The cold electrons at the rear side of the target are then dragged towards the laser region where they are heated. In Figure \ref{fig:shock} j it is possible to distinguish between the population of fast electrons that propagate in the rear side of the target and the bulk of the background electrons that have negative momentum and are being dragged towards the laser due to the electric field that is set up in the plasma (Fig. \ref{fig:shock} n). This leads to the heating of the entire plasma volume and, together with the electron recirculation provides a uniform temperature as can be seen in Figure \ref{fig:shock} k for late times. The fraction of laser light absorbed into the plasma is measured to be 60\% ($\eta = 0.6$). As the uniformly heated plasma expands and a shock is formed, it is also possible to observe that the filamented density structures caused by the laser interaction are smeared out and the shock front becomes relatively uniform. By this time, the laser interaction is finished, and the shock moves at a relatively constant velocity, which is measured to be 0.19 c (Fig. \ref{fig:x_t}) and corresponds to $M_{sh} = 1.7$ for the measured upstream temperature $T_{e\,0} = 1.6$ MeV. The measured Mach number is in good agreement with the theoretical $M_{cr}$ for large $\Gamma$ and $\Theta \sim 1$, $M_{cr} \sim 1.5 - 1.8$ (Fig. \ref{fig:theory}). The shock structure has a strong localized electric field at the shock front, with a measured thickness of $L_{sh}  \sim 4 \lambda_D = 10 c/\omega_0$, where $\lambda_D = \sqrt{k_B T_e/4 \pi n_p e^2}$ is the Debye length, which is much smaller than the mean free path for particle collisions ($L_{sh} \ll \lambda_{e\,i} \sim c/\nu_{e\,i} \sim 2 \times 10^8 \lambda_D$, $\lambda_{i\,i} \sim c_{s\,0}/\nu_{i\,i} \sim 2\times 10^2 \lambda_D$, for $T_e = 1$ MeV, $T_i = 100$ eV, and $n_e = n_i = 10^{21}$ cm$^{-3}$). Ahead of the shock, the TNSA field is approximately constant and in very good agreement with Eq. \eqref{eq:efield} (Fig. \ref{fig:shock} b). The density and field structure is similar to the case of Figure \ref{fig:slabs} e and f, where no laser is used and a denser slab expands into a more rarefied one with an exponentially decreasing density profile.

\begin{figure}[!t]
\begin{center}
\includegraphics[width=1.\columnwidth]{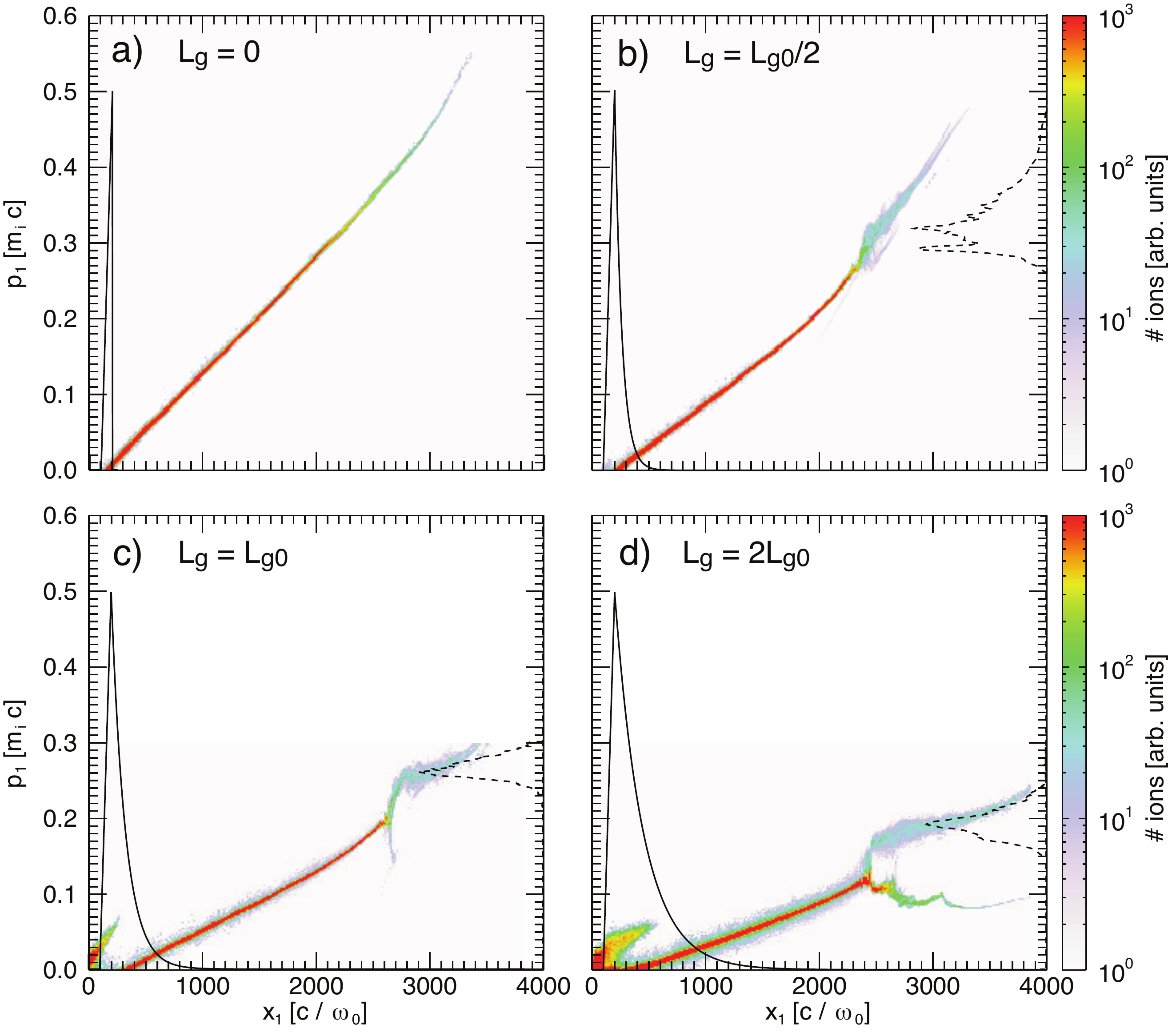}
\caption{\label{fig:optimal} Ion phase-space and spectrum shock accelerated ions (dashed line) for upstream plasmas with different scale lengths: a) $L_g = 0$ (sharp plasma-vacuum transition), b) $L_g = L_{g0}/2$, c) $L_g = L_{g0}$, and d) $L_g = 2 L_{g0}$. The initial density profile is indicated by the solid lines and $L_{g0}$ is given by Eq. \eqref{eq:optimal}.}
\end{center}
\end{figure}

As the shock moves through the upstream expanding plasma it reflects the fraction of the upstream ions which have kinetic energy lower than the electrostatic potential energy of the shock to a velocity of 0.26 c (Fig. \ref{fig:x_t}), which is twice the shock velocity in the upstream frame plus the plasma expansion velocity $v_0$, producing an ion beam with 31 MeV and an energy spread of 12\% (Fig. \ref{fig:shock} t and Fig. \ref{fig:optimal} c). The uniform shock velocity obtained under optimal conditions is crucial to get a uniform velocity in the reflected ions as we can see in Figure \ref{fig:x_t}. The reflected ion beam contains approximately 10\% of the upstream ions, which is consistent with the reflected fraction observed in the interaction of two plasmas with moderate density ratios (Fig. \ref{fig:M_n_r}). The laser to ion beam energy conversion efficiency is $3$ \%. We note that while a high reflection efficiency is desirable in order to accelerate a large number of ions it can have a deleterious effect for the beam quality, since, as previously noted \cite{bib:macchi}, the strong dissipation of the shock will lead to a decrease of its velocity and a chirp in the ion spectrum. Therefore, moderate reflection efficiencies, which are obtained for moderate density ratios/drift velocities, are preferable for the generation of high-quality beams.

We have varied the scale length of the rear side of the plasma in order to validate the optimal conditions for the generation of high-quality beams. We observe that for shorter scale lengths the TNSA fields become dominant leading to a very broad spectrum. For $L_g = L_{g0}/2$ the reflected ions have an average energy of 47 MeV, which is similar to the case of a sharp plasma-vacuum transition (Fig. \ref{fig:optimal} a), but the energy spread was increased to 36\% (Fig. \ref{fig:optimal} b). For a larger scale length ($L_g = 2 L_{g0}$), where it is harder to uniformly heat the entire plasma, the reflected beam has an energy of 17 MeV and an energy spread of 30\% (Fig. \ref{fig:optimal} d). For a very long scale length (uniform profile) the laser cannot heat the entire plasma region and no shock is observed.

We have tested the impact of the laser spot size in the shock formation process and in ion acceleration. Driving a stable shock front and a stable acceleration requires that the shock width (which is close to the laser spot size $W_0$) is large enough such that the plasma, expanding transversely at $c_s$, does not leave the shock width region before the acceleration occurs. Assuming an isothermal expansion, this condition yields $W_0 \gtrsim L_{g\,0}/M_{sh}$, which for $M_{sh} \approx 2$, gives $W_0 \gtrsim 10 \lambda_0$. Simulations performed for the same laser and plasma parameters but using a super-Gaussian transverse laser profile with $W_0 = 16 \lambda_0$, led to the generation of a stable shock and a reflected ion beam with 28 MeV and an energy spread of 9 \%. The energy coupling efficiency from the laser to the ion beam was 2\%. Assuming cylindrical symmetry, the total number of accelerated ions as inferred from the simulation is given by $N_{ions} \sim 10^{10} (W_0 [\mu \mathrm{m}])^2/\lambda_0 [\mu \mathrm{m}]$, where $W_0$ is the laser spot size. This number of ions per bunch is ideal for most applications. For instance, in radiotherapy $\sim 10^8$ ions per bunch are used in multi-shot treatment and $\sim 10^{11}$ ions per bunch in single shot treatment \cite{bib:bulanov,bib:linz}.

The intrinsic ion beam divergence associated with the shock acceleration process can be estimated if we take into account that the velocity of the accelerated ions in the component perpendicular to the shock propagation direction is given by the thermal ion velocity of the upstream plasma and the parallel component is given by approximately twice the shock velocity. The half-angle divergence is then $\theta = \mathrm{tan}^{-1}\left[\frac{1}{2M}\left(\frac{T_i}{T_e}\right)^{1/2}\right]$. For typical moderate Mach numbers ($M \simeq 2$) and electron to ion temperature ratios (in our simulations we observe $T_e/T_i \gtrsim 10$), we expect an half-angle divergence $\lesssim 4.5^{\circ}$, which is consistent with the observed values of $2^{\circ}$ in experiments \cite{bib:haberberger} and $4.1^{\circ}$ in simulations \cite{bib:fiuza2} where a super-Gaussian transverse laser profile has been used. For Gaussian transverse laser profiles the shock front will have a larger curvature which will increase the overall beam divergence, since away from the laser propagation axis the acceleration will occur at an angle.

We note that in 3D the dynamics associated with the laser-plasma interaction in the front of the target (such as self-focusing and filamentation) and with the formation of the space-charge field at the rear side of the target will be different than in 2D. The spot size of a self-focusing laser in a plasma is given by $W = W_0 \sqrt{1-z^2/z_0^2}$, where $z_0 = z_R/\sqrt{P/P_c -1}$ is the typical distance for self-focusing, $z_R = \pi W_0^2/\lambda_0$ is the Rayleigh length, $P$ is the laser power, and $P_c  [\mathrm{GW}] = 17 n_c/n_p$ is the critical power for self-focusing \cite{bib:sprangle}. For typical high-power lasers ($P > 10$ TW) and underdense plasmas ($n_c/n_p \sim 10$), $P/P_c \gg 1$. For a laser spot size capable of driving a stable shock ($W_0 \gtrsim 10 \lambda_0$) the typical self-focusing distance is then $z_0 \gtrsim \frac{130}{\sqrt{P [\mathrm{TW}]}}\lambda_0$. This means that it is important to keep the characteristic rise length of the plasma profile below a few 10 $\lambda_0$ (which is comparable to the optimal scale length of the rear side of the target, $L_{g\,0}$) in order to guarantee that self-focusing does not play an important role. On the rear side of the target, the TNSA field amplitude will be smaller in 3D which should benefit the generation of high-quality shock-accelerated ion beams. 3D PIC simulations of this acceleration process are certainly desirable in order to investigate in detail the role of 3D effects in the acceleration process.

\subsection{Ion energy scaling}

It is of great interest to study the potential of shockwave acceleration to generate ions in the energy range of $100-300$ MeV/a.m.u. required for medical applications \cite{bib:linz}. As the electron temperature increases with increasing laser intensity, it should be possible to generate larger shock velocities and high energy ion beams.

The final ion energy is given by the contribution of both the shock acceleration and the uniform expansion of the upstream plasma. In the relativistic case, the final ion velocity is $v_{ions} = (v_{sh}' + v_{0})/(1+v_{sh}' v_{0}/c^2)$, where $v_{sh}' = (2 M c_{s \,0})/(1+ M^2 c_{s \,0}^2/c^2)$ is the velocity of the reflected ions in the upstream frame and $v_{0}$ is the upstream velocity at the shock acceleration time $t_{acc}$. Taylor expanding for $c_{s \,0}/c \ll 1$, the proton energy for optimal conditions is given by
\begin{eqnarray}
\epsilon_{ions} [\mathrm{MeV}] & \simeq & 2 M_{cr}^2 T_{e \,0} [\mathrm{MeV}] + M_{cr}\frac{t_{acc}}{L_{g\,0}}\frac{(2 T_{e \,0} [\mathrm{MeV}])^{3/2}}{(m_i/m_e)^{1/2}} \nonumber \\
& + & \left[\left(\frac{t_{acc}}{L_{g\,0}}\right)^2 + 4 M_{cr}^4\right]\frac{(T_{e \,0} [\mathrm{MeV}])^{2}}{m_i/m_e}.
\label{eq:scaling}
\end{eqnarray}

\begin{figure}[t!]
\begin{center}
\includegraphics[width=1.\columnwidth]{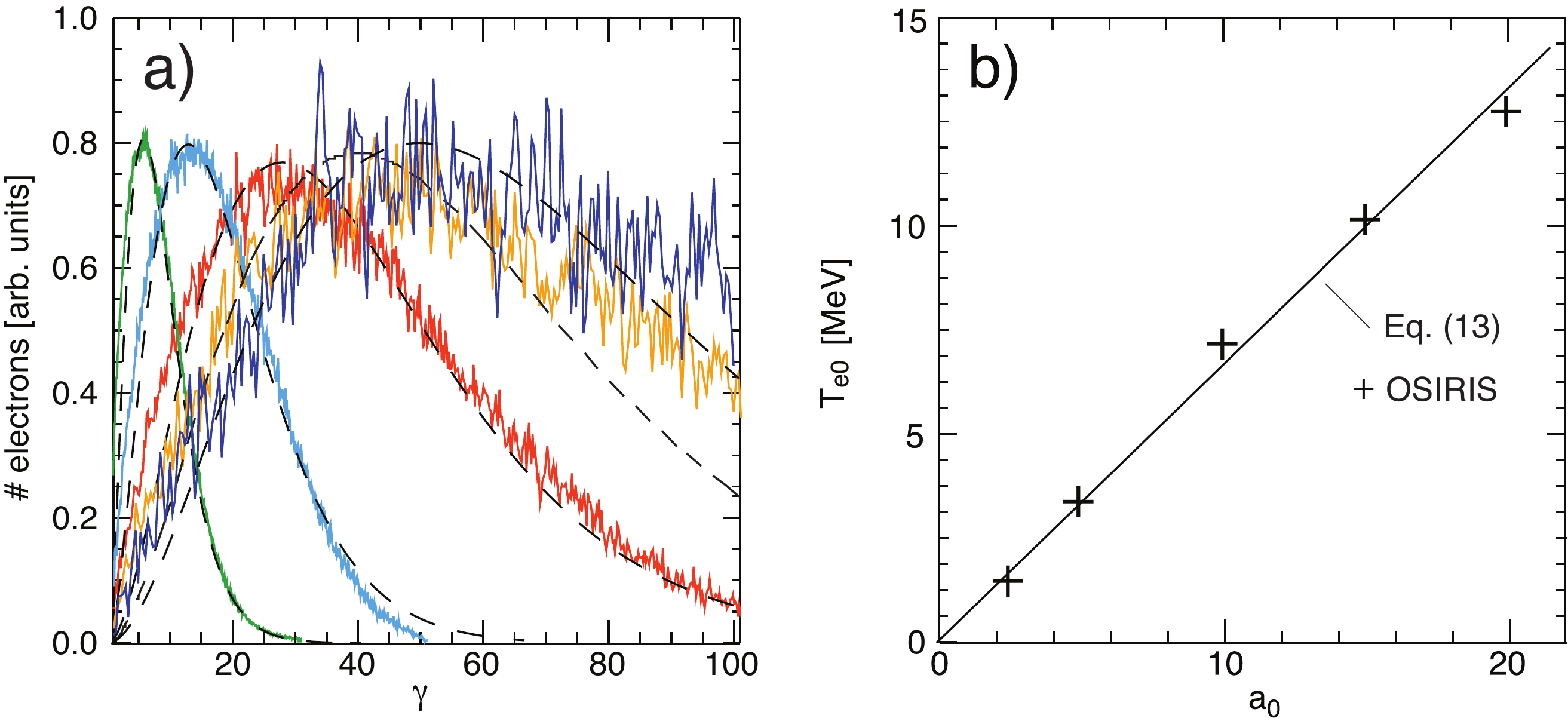}
\caption{\label{fig:te_scaling} Electron distribution for different laser intensities corresponding to $a_0 = 2.5$ (green), 5 (light blue), 10 (red), 15 (orange), and 20 (blue). The distributions are fitted to a 3D relativistic Maxwellian of the form $f(\gamma) = a \gamma^2 e^{-\gamma/\Delta \gamma}$ (dashed lines). b) Scaling of the electron temperature with the laser amplitude $a_0$. The obtained scaling is consistent with Eq. \eqref{eq:temperature} for a laser-electrons energy coupling efficiency $\eta = 0.51$.}
\end{center}
\end{figure}

To investigate the ion energy scaling, 2D simulations have been performed for increasing laser intensities and the same optimal plasma profile. The peak density was increased together with the intensity in order to compensate for the relativistic transparency. The electron temperature is observed to scale linearly with the laser amplitude (Fig. \ref{fig:te_scaling}), which agrees with Eq. \eqref{eq:temperature} for a laser to electron coupling efficiency $\eta = 0.51$ (consistent with our measured laser absorption). 

\begin{figure}[t!]
\begin{center}
\includegraphics[width=1.\columnwidth]{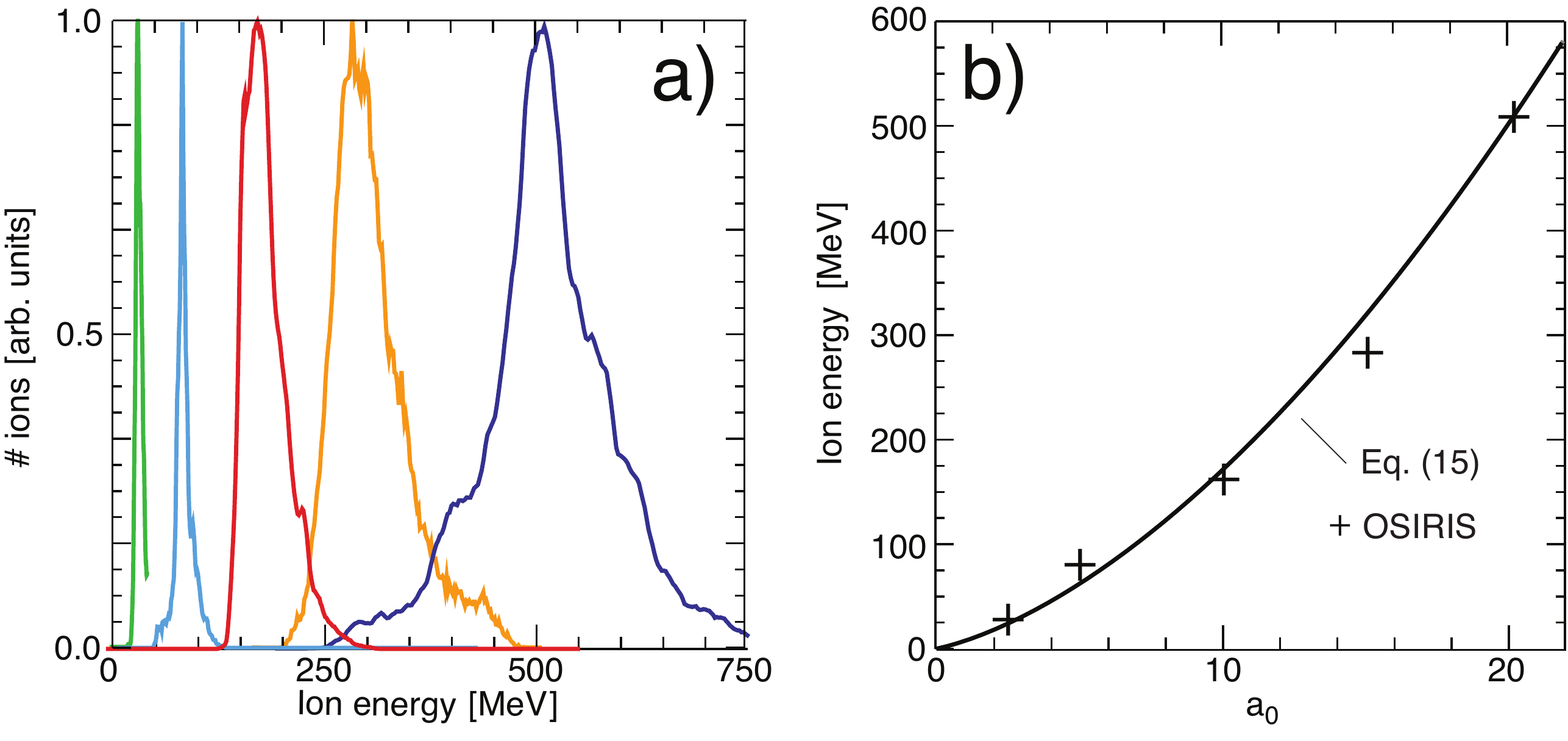}
\caption{\label{fig:ene_scaling} a) Spectrum of shock accelerated ion beams for different laser intensities corresponding to $a_0 = 2.5$ (green), 5 (light blue), 10 (red), 15 (orange), and 20 (blue). b) Scaling of ion energy with the laser amplitude $a_0$. The obtained scaling is consistent with Eq. \eqref{eq:scaling}.}
\end{center}
\end{figure}

For the increased laser intensities increased ion energies are observed up to 512 MeV for $a_0 = 20$ (Fig. \ref{fig:ene_scaling}). The final energy spread varies between 10\% and 25\%. The ion energy scaling with $a_0$ is consistent with Eq. \eqref{eq:scaling} for an acceleration time of $t_{acc} = 5500 \omega_0^{-1}$ (average acceleration time in our simulations). At low intensities the acceleration is dominated by the shock reflection (first and second term of Eq. \eqref{eq:scaling}), but at higher intensities the contribution from the ion expansion (third term of Eq. \eqref{eq:scaling}) also becomes important, leading to a transition form a scaling with $a_0^{3/2}$ to $a_0^2$. This favorable scaling allows for the generation of high quality $\sim 200$ MeV proton beams required for medical applications with a 100 TW class laser system ($a_0 = 10$). 

The generation of 100s MeV ion beams using the proposed scheme can be readily tested experimentally at different facilities where laser systems capable of delivering 100 TW to 1 PW power and pulse durations of 0.5 ps $-$ 1 ps are available. The expanded plasma profiles required ($10\mathrm{s} ~\mu$m scale and $\sim10^{22}$ cm$^{-3}$ peak density) can be obtained from the irradiation of a $\mu$m scale solid foil by a first low-intensity laser and subsequent target expansion. The use of CO$_2$ laser pulses ($\lambda_0 = 10 ~\mu$m) is an alternative possibility \cite{bib:haberberger}, allowing for the use of gas targets where the required plasma profiles, with mm scales and $n_e \sim 10^{19}$ cm$^{-3}$, can be naturally obtained from the ionization of the gas by the laser pre-pulse (or by a train of pulses). The use of gas targets has the important advantage of allowing for high repetition rates in comparison with the conventional solid targets used in ion acceleration experiments.

\section{Conclusions}
\label{sec4}

We have studied the generation of electrostatic shocks in plasma and the use of these shocks to accelerate ions to high energy with low energy spreads. Ion reflection can occur for electrostatic shocks driven by the interaction of plasma regions with large density ratios or moderate relative drift velocities. The energy and number of the reflected ions increases with the density ratio or relative drift velocity. For a finite size plasma, it is important to control the sheath field at the plasma-vacuum interface, and that can be achieved by having an expanded plasma profile with an exponentially decreasing density gradient. In this case, TNSA fields will be approximately uniform and of low amplitude, allowing for a slow expansion of the ions that are then reflected by the shockwave as it reaches the rear side of the plasma.

We have shown that the required conditions to drive strong electrostatic shocks in the laboratory can be obtained by interacting an intense laser with a near critical density tailored plasma. The laser is absorbed near the critical density interface, leading to a local density steepening and heating of the plasma electrons. The fast electrons propagate to the rear side of the target driving an electric field due to the current imbalance that drags the background electrons from the rear side to the laser region. For thin targets, this allows for an efficient heating of the plasma volume. As the heated plasma expands, with an exponentially decreasing density profile, the electrostatic shock can reflect the background ions leading to the generation of a high-energy and high-quality ion beam. The scale length of the plasma profile greatly influences the quality of the accelerated particles. 

It was demonstrated that by increasing the peak density of the plasma in order to compensate for relativistic transparency it is possible to scale this acceleration scheme to the generation of 100s MeV ion beams with current laser systems ($a_0 \sim 10$), which can have an important impact for radiotherapy with compact systems.

\section*{Acknowledgements}
Work supported by the European Research Council (ERC-2010-AdG Grant 267841) and FCT (Portugal) grants PTDC/FIS/111720/2009, SFRH/BD/38952/2007, and SFRH/BPD/65008/2009. Work also performed under the auspices of the U.S. Department of Energy by Lawrence Livermore National Laboratory under Contract DE-AC52-07NA27344 and supported by the LLNL Lawrence Fellowship, and by DOE grant DE-FG02-92-ER40727 and NSF grant PHY-0936266 at UCLA. Simulations were performed at the Jugene supercomputer (Germany) under a PRACE grant, the IST cluster (Lisbon, Portugal), and the Hoffman cluster (UCLA).

\newpage

\end{document}